\documentstyle[aps,epsf,multicol]{revtex}

\title{On Geometric Properties of Passive Random Advection}
\author{Stanislav~A.~Boldyrev
\thanks{E-mail: boldyrev@princeton.edu} 
and Alexander~A.~Schekochihin
\thanks{E-mail: sure@pppl.gov}
\\
{\em Princeton University, P.~O.~Box 451, Princeton, New Jersey, 08543}}
\date{26 July 1999}

\def\secref#1{Sec.~\ref{#1}}
\def\Secref#1{Section~\ref{#1}}
\def\apref#1{Appendix~\ref{#1}}

\def\exref#1{(\ref{#1})}

\def\eqref#1{Eq.~(\ref{#1})}

\def\figref#1{Fig.~\ref{#1}}

\def\virg{,\quad}
\def\et{\quad{\rm and}\quad}
\def\phi{\varphi}

\def\({\left(}
\def\){\right)}
\def\[{\left[}
\def\]{\right]}
\def\<{\left\langle}
\def\>{\right\rangle}
\def\<{\left\langle}
\def\>{\right\rangle}
\def\vl{\left|}
\def\vr{\right|}
\def\fl{\left\{}
\def\fr{\right\}}

\def\vx{{\bf x}}
\def\vy{{\bf y}}
\def\dt{{\partial_t}}
\def\d{\partial}

\def\bea{\begin{eqnarray}}
\def\eea{\end{eqnarray}}

\begin{document}
\input psfig.sty
\maketitle

\begin{abstract}
We study geometric properties of a random Gaussian short-time correlated 
velocity field by considering statistics of a passively advected metric 
tensor. That describes universal properties of fluctuations of tensor
objects frozen into the fluid and passively advected by it.  
The problem of one-point statistics of co- and contravariant tensors 
is solved exactly, provided the advected fields do not 
reach dissipative scales, which would break the symmetry of the problem. 
Asymptotic ($t\to\infty$) duality of the problem is established, 
which in the three-dimensional case relates the probabilities 
of the volume deformations into ``tubes'' and into ``sheets''.\\

\end{abstract}

\begin{multicols}{2}

\section{Introduction}

A problem of passive advection in a turbulent medium attracts considerable 
attention as a solvable model of turbulence. Exact solutions can be found 
in a simplified case, when the velocity field is chosen to be a random, 
short-time correlated Gaussian process. Statistics of density, concentration, 
passive vectors advected by such a field were investigated by many authors 
(see,~e.~g.,~\cite{Kraichnan-1,Kraichnan-2,Shraiman-1,Chertkov-3,Bernard_Gawedzki_Kupiainen,Balkovsky_Fouxon}), 
where intermittent nature of the fluctuations, non-trivial scalings of 
structure functions and anomalous role of the dissipation were discovered. 
All these features are very common in the general picture of turbulence and, 
therefore, the problem of passive advection can serve as a model for
developing corresponding analytical tools.

In the present paper we consider passive advection (in the Lie sense) 
of a second-rank covariant tensor in $d$-dimensional space. Though our 
master equation for the probability density function~(PDF) of the tensor 
(\eqref{eqn_Z} below) is very general, we concentrate mainly on statistics 
of a symmetric (metric) tensor~$g_{ij}$. One-point statistics of any tensor 
object frozen into the fluid can be related to statistics of such a tensor. 
We do not impose any restrictions (such as incompressibility) 
on the velocity field, and, therefore, statistics in both Eulerian 
and Lagrangian frames are studied. 
Also, we are only interested in the ``initial stage'' of the advection, 
when the advected field does not reach dissipative scales. This allows us 
to explore the symmetries of the problem, which are broken when dissipation 
is included. 

We show that the probability-density function of the eigenvalues 
of the metric is governed by a $d$-particle Hamiltonian that can be 
split into two non-interacting parts. Its {\em non-universal} part describes 
the motion of the center of mass (the determinant~$g$ of the metric) 
and can be separated from the motion relative to the center of mass, 
i.~e.~dynamics of the metric's eigenvalues normalized to their geometrical 
mean,~$\lambda_i/g^{1/d}$. The Hamiltonian of the latter motion is 
of the Calogero-Sutherland type, 
remains the same in both Lagrangian and Eulerian frames of reference, 
and therefore describes the {\em universal} properties of the advection. 
These properties are dictated by the symmetry of the problem. The exact 
integrability of the Calogero-Sutherland Hamiltonian is known to be related 
to~$SL(d)$ symmetry: the Hamiltonian can be represented 
as a quadratic polynomial in terms of the generators of the corresponding 
algebra~\cite{Bernard_Gawedzki_Kupiainen,Shraiman_Siggia,Ruehl_Turbiner}. 
The eigenfunctions of this Hamiltonian are the so-called Jack polynomials, 
which are symmetric homogeneous functions of the eigenvalues. 
This allows us to find exactly 
all moments~$\langle\hat T^m\rangle$ of any tensor~$\hat T$ advected by the 
fluid. Indeed, calculating any such moment reduces to averaging expressions 
of the type~${\rm Tr}^k({\hat g}^n)$, which are symmetric polynomials 
in terms of the metric's eigenvalues, and can therefore be expanded 
in Jack polynomials of degree~$nk$. We illustrate this method 
by calculating exactly all moments of passively advected vectors 
and covectors, in particular, of the magnetic field in kinematic 
r\'egime and of the passive-scalar gradient. We also demonstrate 
how this approach works in the general case of a passively advected 
tensor of any rank. 

Calculating the moments requires knowing the statistics of the 
metric~$\hat g$ with special initial conditions,~$g_{ij}(t=0)=\delta_{ij}$. 
However, it is also interesting to consider 
the evolution of the~PDF of the symmetric tensor~$g_{ij}$ subject to 
arbitrary initial conditions. In this context, we show that a beautiful 
dual picture exists: the time-dependent~PDF  of the tensor becomes 
asymptotically ($t\to\infty$) invariant under the inversion of 
the eigenvalues with respect to their geometrical mean. For example, 
in three dimensions, that means that if a magnetic field advected by 
ideally conducting fluid develops flux tubes, it must develop magnetic 
sheets with the same probability. 

The paper is organized as follows. In \Secref{master}, we derive the master 
equation for the~PDF of the metric's eigenvalues, and analyze the symmetry 
properties of this~PDF. In \Secref{EL_PDF}, we present a simple method of 
transforming the~PDF between Eulerian and Lagrangian frames, which is 
important in the case of a compressible velocity field. \Secref{solutions} 
discusses general properties of solutions for the~PDF in two- and 
three-dimensional cases. In \Secref{passive_vector}, we show how 
the symmetry of the problem allows to calculate all the moments of passively 
advected tensors. The paper is written in a self-contained manner, 
all the necessary definitions and derivations are summarized in 
the Appendices.

\section{Master Equation}
\label{master}

A covariant second-rank tensor field $\phi_{ij}(t,\vx)$ passively advected 
by the velocity field~$\xi^k(t,\vx)$ evolves according to the 
following equation:
\bea
\label{dynamic_g}
\dt \phi_{ij} + 
\xi^k\phi_{ij,k} + \xi^k_{,i}\phi_{kj} + \xi^k_{,j}\phi_{ik} = 0,
\eea
where $\xi^k_{,i}=\d\xi^k/\d x^i$, 
and~$\phi_{ij,k}=\d\phi_{ij}/\d x^k$. Let $\xi^i(t,\vx)$ 
be a Markovian Gaussian field: 
\bea
\nonumber
&\<\xi^i(t,\vx)\xi^j(t',\vx')\> = 
\kappa^{ij}(\vx-\vx')\delta(t-t'),\\
\label{correlator_xi}
&\kappa^{ij}(\vy) \simeq \kappa_0\delta^{ij} 
- \kappa_2\(y^2\delta^{ij} + 2ay^i y^j\)\virg y\to0,
\eea 
where~$a$ is the compressibility parameter, and~$\kappa_2=1$ for 
simplicity. Here~$a$ can vary between~$-1/(d+1)$ for the incompressible flow 
and~$1$ for the fully compressible flow. 

In order to determine the statistics of the tensor, we follow a standard 
procedure~\cite{Polyakov,Boldyrev} and introduce the characteristic 
function of~$\hat\phi(t,\vx)$:
\bea
Z(t,\hat\sigma) = \<\exp\fl\sigma^{ij}\phi_{ij}(t,\vx)\fr\>. 
\eea
This function is a Fourier transform of the~PDF of the matrix 
elements~$\phi_{ij}$. 
Clearly, $Z$ is independent of~$\vx$ due to spacial homogeneity.
We find that~$Z$ satisfies
\bea
\nonumber
\dt Z & =& -\[1+a(d+1)\]\sigma^{ij}{\d Z\over\d\sigma^{ij}} + 
2a\sigma^{ij}{\d\over\d\sigma^{ij}}\sigma^{mn}{\d Z\over\d\sigma^{mn}} \\
\nonumber
&+&\,{1\over2}\(\sigma^{ij}{\d\over\d\sigma^{kj}} + 
\sigma^{ji}{\d\over\d\sigma^{jk}}\)
\(\sigma^{il}{\d\over\d\sigma^{kl}} + 
\sigma^{li}{\d\over\d\sigma^{lk}} \right. \\
\label{eqn_Z} 
&+&\left.a\sigma^{kl}{\d\over\d\sigma^{il}} + 
a\sigma^{lk}{\d\over\d\sigma^{li}}\)\,Z,
\eea
where~$d$ is the dimensionality of space.
This equation was derived by taking the time derivative of~$Z$, 
using \eqref{dynamic_g}, and splitting Gaussian averages. 
We obtain the equation for the probability density function 
of~$\hat\phi$ by Fourier-transforming \exref{eqn_Z}:
\bea
P(\hat\phi) = \int\exp\fl-i\sigma^{ij}\phi_{ij}\fr Z(\hat\sigma)
\prod_{m,n}{\rm d}\sigma^{mn}.
\label{pdf}
\eea

The original equation (\ref{dynamic_g}) preserves symmetry properties of the 
tensor~$\phi_{ij}$, which means that we may restrict our consideration 
either to advection of symmetric or antisymmetric tensors. Both reductions 
can be done in a similar fashion. For our present purposes we only consider 
fluctuations of a symmetric covariant tensor. 
The corresponding results for a contravariant tensor are summarized 
in~\apref{ap_contra}. We will use both (co- and contravariant) 
pictures when discussing statistics of passive vectors 
in \Secref{passive_vector}.

In the symmetric case, the PDF~\exref{pdf} can be factorized as follows:
\bea
P(\hat\phi) = \tilde P(\hat g) \prod_{m<n}\delta(\phi_{mn}-\phi_{nm}),  
\eea
where~$\hat g$ is the symmetric part of~$\hat\phi$. One may think of 
the tensor~$\hat g$ as of a metric associated with the medium. 
Due to spacial isotropy, $\tilde P$~depends only on the eigenvalues 
$\lambda_1,\dots,\lambda_d$ of $\hat g$. After rather cumbersome but 
essentially simple calculations, we establish the following master 
equation for the PDF of the eigenvalues of the~metric:
\bea
\nonumber
\dt P& =& 2\,(2a+1)\sum_i\lambda_i^2{\d^2 P\over\d\lambda_i^2} + 
2a\sum_{i\neq j}\lambda_i\lambda_j{\d^2 P\over\d\lambda_i\d\lambda_j} \\
\nonumber
&+& \[3d+4+2a(d^2+3d+3)\]\sum_i\lambda_i{\d P\over\d\lambda_i} \\
\nonumber  
&{}&+\,(a+1)\sum_{i\neq j}{\lambda_i\lambda_j\over\lambda_i-\lambda_j} 
\({\d P\over\d\lambda_i}-{\d P\over\d\lambda_j}\) \\
\label{master_eq}
&{}&+\,{1\over2}\,d\,(d+1)(d+2)\[1+a(d+1)\]P
\eea
(from here on the overtildes are dropped).
Among the solutions of this equation, those corresponding to the PDF 
must  be 
non-negative, finite, and normalizable. The normalization is as 
follows \cite{Zinn-Justin}:
\bea
\int {\rm d}\lambda_1\dots {\rm d}\lambda_d P(\lambda_1,\dots,\lambda_d) 
\prod_{i<j}\vl\lambda_i-\lambda_j\vr = 1.
\eea
Clearly, the original stochastic  equation~\exref{dynamic_g} 
preserves the signature of the metric. We will restrict ourselves to 
the case of all positive~$\lambda$'s.
Since there is no means of distinguishing between 
different orderings of the eigenvalues, the~PDF must be a symmetric 
function with respect to all permutations of~$\lambda_1,\dots,\lambda_d$. 

We should now notice that in logarithmic variables $z_i=\log(\lambda_i)$, 
the master equation~\exref{master_eq} describes the dynamics of 
$d$~pair-wise interacting particles on the line. Furthermore, 
we can consider these dynamics in the reference frame associated 
with the center of mass of the particles~$z={1\over d}\sum z_i$. 
Denoting the coordinates of the particles in this frame~$\zeta_i=z_i-z$, 
and noticing that~$\det(\hat g)=g=\exp(zd)$, we find that~$P$ now satisfies 
\bea
\nonumber
&{}&\dt P = d\,[1+a(d+1)]\fl 2g^2{\d^2 P\over\d g^2} + 
(2d+5)\,g\,{\d P\over\d g}\right. \\ 
\nonumber
&+& \left.{1\over2}(d+1)(d+2)\,P\fr + 2(1+a)\[-{1\over d}
\sum_{i,j}^d{\d^2 P\over{\d\zeta_i\d\zeta_j}}\right. \\
\label{master_eq_gzeta}
&+& \left. \sum_{i=1}^d{\d^2 P\over\d\zeta_i^2} 
+ {1\over2}\sum_{i<j}^d{1\over\tanh{1\over2}(\zeta_i-\zeta_j)} 
\({\d P\over\d\zeta_i} - {\d P\over\d\zeta_j}\)\],
\eea
where the $d$~variables~$\zeta_1,\dots,\zeta_d$ are not independent, 
$\sum\zeta_i=0$. 
The Hamiltonian remaining after the dynamics of the center of mass are 
separated, is translationally invariant, therefore the total momentum 
of the particles~$\sum(\d P/\d\zeta_i)$ is conserved. 
The normalization rule now is:
\bea
\nonumber
\int{\rm d}\zeta_1\cdots{\rm d}\zeta_d\,\,&{}&\delta\(\zeta_1+\cdots+\zeta_d\) 
\vl J(\zeta)\vr 
\int{\rm d}g\,g^{(d-1)/2}P = 1,  \\
\label{Jacobian_gzeta}
J(\zeta) &{}&= {2^{d(d-1)/2}\over d}
\prod\limits_{i<j}^d\sinh{1\over2}\(\zeta_i-\zeta_j\), 
\eea
where by~$\zeta$ we denote the set $\{\zeta_1,\dots,\zeta_d\}$.
The operator in the square brackets is a Sutherland Hamiltonian~$\tilde H_S$, 
which is exactly solvable~(see, 
e.g.,~\cite{Ruehl_Turbiner,Sutherland,Olshanetsky_Perelomov}; 
this Hamiltonian appeared in a similar context 
in~\cite{Bernard_Gawedzki_Kupiainen,Shraiman_Siggia}). The 
Hamiltonian~$\tilde H_S$ is the same for  co- and contravariant tensors, 
and in both Eulerian and Lagrangian frames.

It is important that~$\tilde H_S$ 
is {\em self-adjoint} with respect to the measure~\exref{Jacobian_gzeta}.
Its eigenfunctions are the so-called Jack polynomials, that are homogeneous 
polynomials in~$\exp(\zeta_i)$ and are symmetric with respect to all 
permutations of~$\zeta_i$. Their construction is discussed 
in \apref{Jack_polynomials}. We will use particular eigenfunctions of this 
operator in~\secref{passive_vector}. 

We see that if~$P$ is initially chosen in a factorized form, 
$P=P_1(g)P_2(\zeta_1,\dots,\zeta_d)$, it will remain so factorized 
at all times. Thus, the statistics of~$g$ are independent of the statistics 
of the~$\zeta$'s at all times if they are initially independent. 
In particular, this property of~\eqref{master_eq_gzeta} allows us 
to consider separately the~PDFs for the determinant of the metric 
and for the logarithmic quantities~$\zeta_i=\log(\lambda_i g^{-1/d})$:
\bea
\nonumber
 S(t,g)& = & \int{\rm d}^d\zeta\,|J(\zeta)|P(g,\zeta),\qquad\,\, \\
 F(t,\zeta)& = &\int{\rm d}g\,g^{(d-1)/2}P(g,\zeta).
\eea

An additional symmetry emerges in this context:~\exref{master_eq} 
and~\exref{Jacobian_gzeta} remain 
invariant if the coordinates~$z_i$ of all particles are simultaneously 
{\em reflected} with respect to their center of mass. Such reflection 
leaves the center of mass intact and reverses the signs of all~$\zeta_i$, 
i.~e.~transforms all~$\lambda_i$ into $g^{2/d}/\lambda_i$. 
The origin of this symmetry can be understood if we notice that 
the master equations for the~PDFs of the dimensionless 
quantities~$G_{ik}=g^{-1/d}g_{ik}$ and~$G^{ik}=g^{1/d}g^{ik}$ are the same, 
although the initial stochastic equations are different.  
This symmetry leads to nontrivial results for~$d\ge3$, and will 
be considered in~\Secref{solutions}.

\section{Eulerian and Lagrangian PDF's}
\label{EL_PDF}

The equation for the metric-determinant PDF~$S(t,g)$ follows 
from~\eqref{master_eq_gzeta}:
\bea
\label{eq_S_E}
\dt S = 2g^2{\d^2 S\over\d g^2} + (2d+5)\,g\,{\d S\over\d g} 
+ {1\over2}(d+1)(d+2)S, 
\eea 
where we have rescaled time by the factor of $\gamma=d\,[1+a(d+1)]$. 
This factor is always non-negative and vanishes 
if the velocity field is incompressible, $a = -1/(d+1)$, 
in which case any time-independent function $S(g)$ is a solution. 
Note that the right-hand side of~\eqref{eq_S_E} becomes 
a full derivative when multiplied by the Jacobian $g^{(d-1)/2}$.
The solution of this equation is a log-normal distribution:
\bea
S(t,g) = {g^{-(d+1)/2}\over\sqrt{8\pi\gamma t}} 
\exp\fl-{\bigl(\log(g)+\gamma t\bigr)^2\over8\gamma t}\fr,
\label{S_Euler}
\eea
where we took the initial distribution in the form $S(0,g)=\delta(g-1)$.

This result can be simply understood if we note that the 
determinant~$g$ obeys the same equation as~$\rho^2$, 
the squared density of the medium.
The density satisfies the continuity equation, which can be 
written in logarithmic~form:
\bea
\label{dynamic_rho}
\dt\log\rho + \xi^k\d_k\log\rho + \xi^k_{,k} = 0.
\eea
Since the time increments of~$\xi^k$ are independent identically 
distributed random variables, the Central Limit Theorem implies  
the normal distribution of~$\log\rho$. Indeed, either 
from~\eqref{eq_S_E} or directly from~\eqref{dynamic_rho}, 
one can easily establish 
that the density~PDF $R(t,\rho)=2\rho^d S(t,\rho^2)$ satisfies 
$\dt R = (\gamma/2)\,(\rho^2 R)''$.

So far, we have worked in the Eulerian frame, considering statistics 
at an arbitrary fixed point~${\bf x}$. Now we show 
how the one-point joint Eulerian and Lagrangian~PDF's are related. 
Let us assume that initially Lagrangian particles are uniformly distributed 
in space. We denote the Eulerian~PDF~$P_E(\rho,\zeta;t,\vx)$,
the Lagrangian~PDF~$P_L(\rho,\zeta;t,\vy)$, where~$\vy$ is the 
Lagrangian label (initial coordinate of the Lagrangian particle), 
and~$\rho=|\det(\d\vy/\d\vx)|$ (the density of the medium). 
The relation between~$P_E$ and~$P_L$ can be established
from the following:
\bea
\nonumber
P_E(\rho,\zeta;t,\vx) = \bigl<\delta\(\rho-\rho(t,\vx)\)
&{}&\delta\(\zeta-\zeta(t,\vx)\)\bigr>= \\
\label{EL_id}
\int{{\rm d}\vy\over\rho}\bigl<\delta\(\vx-\vx(t,\vy)\)
\delta\(\rho-\rho(t,\vy)\)&{}&\delta\(\zeta-\zeta(t,\vy)\)\bigr>.
\eea
Since the one-point~PDF~$P_E(\rho,\zeta;t,\vx)$ is independent of position 
(due to spacial homogeneity), we can integrate~\exref{EL_id} with respect 
to~$\vx$. Also noting that the one-point~PDF~$P_L(\rho,\zeta;t,\vy)$ is 
independent of~$\vy$, we get:
\bea
\label{EL_transform}
P_E(\rho,\zeta) = {1\over\rho}\,P_L(\rho,\zeta).
\eea 
Transformation to the Lagrangian frame can also be performed on the level 
of the original stochastic equations such as~\exref{dynamic_rho} 
with the aid of the stochastic calculus 
(see, e.~g.,~\cite{Zinn-Justin,Oksendal}).

In our considerations, if we choose 
intially~$S(0,g)\propto\delta(g-\rho_0^2)$, we may 
substitute~$\rho=\sqrt{g}$ in formula~\exref{EL_transform}.
We see therefore that only the~PDF of~$g$ is affected 
by the transformation between Eulerian and Lagrangian frames.
The Lagrangian version of~$S(g)$ is:
\bea
S(t,g) = {g^{-(d+1)/2}\over\sqrt{8\pi\gamma t}} 
\exp\fl-{\bigl(\log(g)-\gamma t\bigr)^2\over8\gamma t}\fr.
\label{S_Lagrange}
\eea
Analogous results for the contravariant case are presented 
in~\apref{ap_contra}.

The log-normal statistics such as~\exref{S_Euler} and~\exref{S_Lagrange} 
are a signature of this problem, and they will also be present 
for fluctuations of the eigenvalue ratios in asymptotically-free r\'egimes, 
i.~e.~where different ratios do not interact with each 
other~\cite{Kraichnan-1,Kraichnan-2,Shraiman-1,Chertkov-3,Bernard_Gawedzki_Kupiainen,Balkovsky_Fouxon}.

\section{PDF's of Eigenvalue Ratios in Two and Three Dimensions}
\label{solutions}

We saw in the previous section that~$F(t,\zeta)$, 
the~PDF of the ratios~$\lambda_i/g^{1/d}$, would remain the same 
in both Eulerian and Lagrangian frames. In this section we analyse 
the equations for these~PDF's in two- and three-dimensional cases. 
Having in mind numerical simulations, we will write 
these equations using~$d-1$ {\em independent} variables. In the general 
case such reduction is done in \apref{PDF_independent_variables}.

Let us start with the two-dimensional case. It is now convenient to integrate 
the $\delta$-function in~\exref{Jacobian_gzeta} and work with 
the logarithm of the eigenvalue ratio as a new 
variable:~$x={1\over2}\log(\lambda_1/\lambda_2)={1\over2}(\zeta_1-\zeta_2)$.
The equation for~$F(t,x)$ then becomes 
\bea
\label{eq_F2_log}
\dt F = (1+a)\[F_{xx}''+{1\over\tanh(x)}\,F_x'\].
\eea
As expected, the rhs of~\eqref{eq_F2_log} becomes a full derivative 
when multiplied by the Jacobian $J(x)=2\sinh(x)$. Note that 
the differential operator in the right-hand side of~\eqref{eq_F2_log} 
becomes a Legendre operator under the change of 
variables~${\tilde x}=\cosh(x)$. This property is a consequence of 
integrability of the initial Hamiltonian~$H_S$~(\eqref{master_eq_gzeta}), 
and will be of use in~\secref{passive_vector} when we calculate 
the moments of passive vectors.

The nature of the solution can be easily understood if we first consider 
only the advective term $F'_x/\tanh(x)$. The characteristic 
of~\eqref{eq_F2_log} then satisfies~$\dot x = 1/\tanh(x)$, which implies 
that $F$~is advected to regions where~$|x|\gg1$, and, for~$t\to\infty$, 
the asymptotic solution can be found from~\exref{eq_F2_log} by 
approximating $\tanh(x)\approx1$. 
The asymptotic is log-normal as expected.

Note that the reflection symmetry~$x\to-x$ of~\eqref{eq_F2_log} is 
just a consequence of the previously mentioned general 
symmetry~$\lambda_1\leftrightarrow\lambda_2$, and does not add 
anything new. The function~$F$ must be {\em initially} chosen in such 
symmetric form. This is not so in the three-dimensional case that 
we now consider in more detail.

In three dimensions, integrating the $\delta$-function 
in~\exref{Jacobian_gzeta} as before and introducing new 
variables,~$x={1\over2}\log(\lambda_1/\lambda_3)$ 
and~$y={1\over2}\log(\lambda_2/\lambda_3)$,  
we obtain the equation for~$F(t,x,y)$:
\bea
&{}&\dt F = (1+a)\biggl\{ F_{xx}''+F_{xy}''+F_{yy}''\biggr. \nonumber \\
\nonumber
&+& \left( {1\over\tanh(x)}+{\sinh(x)\over2\sinh(y)\sinh(x-y)}\right)\,F_x' \\ 
\label{eq_F3_log}
&+&\left.\left( {1\over\tanh(y)}+{\sinh(y)\over2\sinh(x)\sinh(y-x)}\right) \,F_y'\fr.
\eea
The normalization Jacobian for this PDF 
is~$J(x,y)={32\over3}\sinh(x)\sinh(y)\sinh(x-y)$.

The symmetry with respect to all permutations of 
eigenvalues~$\lambda_1,\lambda_2,\lambda_3$, leads to the following 
two symmetries of the solutions of~\eqref{eq_F3_log}:
\bea
\label{sym_permut_3D}
x\to-x \virg y\to y-x; \et x\leftrightarrow y.
\eea
\eqref{eq_F3_log} posesses another (reflection) symmetry as well:
\bea
\label{sym_refl_3D}
x\to-x \virg y\to-y,
\eea
which corresponds to the inversion 
of~$\lambda_1/\lambda_3$,~$\lambda_2/\lambda_3$, and 
does not follow from~\exref{sym_permut_3D}. Therefore a general 
initial distribution should contain both symmetric,~$F^s$, and 
antisymmetric,~$F^a$, parts with respect to this reflection. 
The symmetries~\exref{sym_permut_3D} act as 
reflections~\exref{sym_refl_3D} on the points 
of the plane located on the lines~$y=2x$,~$y=x/2$, and~$y=-x$; hence 
the antisymmetric part of the~PDF $F^a$~must vanish on these lines.  

Characterictic trajectories of \eqref{eq_F3_log} are presented 
in~\figref{fig1}. The lines~$y=\pm x$, $y=2x$,~$y=x/2$, 
$x=0$,~and~$y=0$ are combined in groups that are transformed by the
symmetries~\exref{sym_permut_3D} independently. Those groups correspond 
to sheet, tube, and strip volume deformations as shown. 
Let us concentrate our attention on the sector~$x\ge0$,~$y\le0$. 
Due to the symmetries~\exref{sym_permut_3D} and~\exref{sym_refl_3D}, this 
allows us to understand the behavior of the~PDF in the entire 
plane~$(x,y)$. Considering the characteristic trajectories 
(they advect~$F$ towards the line~$y=-x$ from both sides), or the flux 
of the conserved function~$F(x,y)|J(x,y)|$ (calculated on the 
line~$y=-x$, it is found to be directed from the semisector with 
positive~$F^a$ to that with negative~$F^a$), one can show that the 
antisymmetric part of the~PDF decays with time. The symmetry of 
the solution with respect to the sheet and tube configurations 
thus emerges asymptotically as $t\to\infty$.

{\columnwidth=3in
\begin{figure} [tbp]
\centerline{\psfig{file=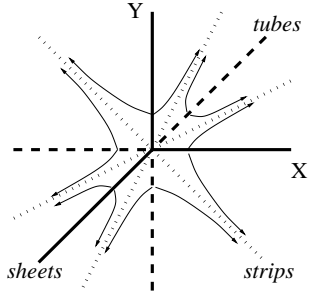,width=4cm}}
\vskip5mm
\caption{Characteristic trajectories of \eqref{eq_F3_log}. Solid lines 
correspond to sheet configurations, dashed lines to tubes, and dotted 
lines to strips.}
\label{fig1}
\end{figure}
}

Numerical simulations performed for various initial distributions 
concentrated in the region~$|x|\le1$,~$|y|\le1$, confirm that the~PDF 
becomes symmetrized very fast, at times~$t\sim 1$. 
In the region~$|x|\gg1$,~$|y|\gg1$, far from 
the lines~$y=x$,~$y=0$, and~$x=0$, the long-time ($t\gg1$) asymptotic is 
log-normal. This asymptotic can be easily obtained from~\eqref{eq_F3_log}. 

\section{Passive Vectors}
\label{passive_vector}

In this section we apply the developed formalism to passively advected 
vectors. Consider the evolution of the coordinates of a particle advected 
by the fluid: $x^i=x^i(t,\vy)$,~where $y^i$~is the initial position of 
the particle, i.~e.~$x^i(0,\vy)=y^i$. An infinitesimal contravariant 
vector~$a^i$ changes under such coordinate transformation 
as follows:~$a^i(t,\vx)= (\d x^i/\d y^k)\,a^k_0(\vy)$.
In order to find the mean of any object constructed out of~$a^i$, 
we have to average it with respect to the initial distribution 
of~$a^i_0(\vy)$ and with respect to all realizations of the random 
velocity field~$\xi^k$. The latter averaging can be done via the~PDF's 
for co-~and contravariant (metric) tensors. 
Let us assume that the initial
distribution of the vector~$a^i$ is Gaussian, isotropic, and 
independent of~$\vy$: $\bigl<a^i_0 a^j_0\bigr>=\delta^{ij}$. 
As an example, consider the moments~$A_n=\bigl<|{\bf a}|^2\bigr>$:
\bea
A_n = \bigl<\({\bf a}_0\cdot{\hat g}\cdot{\bf a}_0\)^n\bigr>, 
\label{moments_contra}
\eea
where~${\hat g}$ is the {\em contravariant} tensor advected by the 
fluid and with the initial condition~$g^{ij}(0,\vy)=\delta^{ij}$. 
The distribution of this tensor can be found in the same way as 
that of the covariant tensor, and is discussed in~\apref{ap_contra}. 
Moments of a {\em covariant} vector~$a_i$ can be found using exactly 
the same formula~\exref{moments_contra}, with~${\hat g}$ now the 
{\em covariant} tensor. 

To simplify the formula~\exref{moments_contra}, we note that 
the eigenvalues of the matrix~$\hat g$ can be expressed 
as~$\lambda_i=g^{1/d}\exp(\zeta_i)$.
Therefore, for all~$n$, $g^{-n/d}\,{\rm Tr}\,({\hat g^n})$~depend 
only on~$\zeta_i$, and are independent of the determinant~$g$.
Since the initial distribution of~$g$ is~$S(g)=\delta(g-1)$, 
we can average powers of~$g$ independently and obtain:
\bea
\nonumber
A_n &=& f_d(n,t)\,\bigl<g^{n/d}\bigr>,\\
f_d(n,t) &=& \bigl<g^{-n/d}\,\({\bf a}_0\cdot{\hat g}\cdot{\bf a}_0\)^n\bigl>,  
\label{moments_universal}
\eea
where the functions~$f_d(n,t)$ do not depend on the statistics of 
the determinant, and are, therefore, {\em universal}. These functions are 
the same in the co- and contravariant cases, and in both Eulerian 
and Lagrangian frames. The only parts of the moments~$A_n$ that are 
non-universal are the averages of the determinant. These averages can 
be calculated exactly using formulas~~\exref{S_Euler},~\exref{S_Lagrange}, 
and~\exref{S_contra}:
\bea
\label{E_co}
\<g^s\>_E^{\rm co}& =& \<g^s\>_L^{\rm contra} = e^{s(2s-1)\gamma t},\\
\label{E_contra}
\<g^s\>_L^{\rm co}& =& \<g^s\>_E^{\rm contra} = e^{s(2s+1)\gamma t}.
\eea

The universal functions~$f_d(n,t)$ can, in fact, be easily calculated 
directly~(cf.~\cite{Chertkov-3,Bernard_Gawedzki_Kupiainen,Fouxon,Chertkov_etal_dynamo}),
if one starts from the equation for advection 
of a passive vector~$a^i(t,\vx)$:
\bea
\dt a^i + \xi^k a^i_{,k} - \xi^i_{,k} a^k = 0,
\eea
where statistics of~$\xi^i(t,\vx)$ are given by~\exref{correlator_xi}. 
However, for methodical purposes, we prefer to rederive this result 
using the technique of Jack polynomials. While it is also quite simple, 
it illustrates the general method that can be applied to finding moments 
of {\em any} passively advected tensor. At the end of this section, 
we show,~e.~g., how moments of a bilinear form~$a^i b^k$ can be caclulated.

Formula~\exref{moments_universal} can be further simplified 
if we do the average with respect to the distribution of~$a_0^i$.
Introducing the generating function
\bea
Z(\beta) = \<\exp\fl\beta\,g^{-1/d}\,
\({\bf a}_0\cdot{\hat g}\cdot{\bf a}_0\)\fr\>, 
\eea
we represent~$f_d$ as follows:
\bea
\label{f_functions}
f_d(n,t) = \[{\d^n Z(\beta)\over\d\beta^n}\]_{\beta=0}.
\eea
The Gaussian average with respect to the initial distribution of 
the vector can now be easily done, resulting~in
\bea
Z(\beta) = \<\prod\limits_{i=1}^d
\Bigl(1 - \beta\exp(\zeta_i)\Bigr)^{-1/2}\>,
\label{generating_function}
\eea
where the remaining averaging is with respect to the statistics of~$\zeta_i$. 
The~PDF of the~$\zeta$'s is~$F(\zeta)|J(\zeta)|\delta(\sum\zeta_i)$ 
with the initial condition~$\delta(\zeta_1)\cdots\delta(\zeta_d)$.
It is important that the function that is being averaged 
in~\exref{generating_function} is the generating function for a particular 
class of Jack polynomials, that are eigenfunctions of the self-adjoint 
Sutherland operator~$H_S$ in~\exref{master_eq_gzeta}. 
Therefore, all functions~\exref{f_functions} can be found exactly 
in the general case. The appropriate calculation is carried out 
in~\apref{Jack_polynomials}. The answer is:
\bea
f_d(n,t)=\({d\over 2}\)_{n}
\exp\fl{d-1\over d}\,n(2n+d)(1+a)t\fr,
\label{f_function_general}
\eea
where we denote:~$(d/2)_n=(d/2)(d/2+1)\cdots(d/2+n-1)$.

In the two-dimensional case the corresponding result can be obtained 
in a rather simple manner, which nevertheless illustrates the main idea 
of the general derivation. In order to do this, we notice that 
the generating function~$Z(\beta)$, expressed in the two-dimensional 
case in terms of~$x={1\over2}(\zeta_1-\zeta_2)$ (see~\secref{solutions}), 
coincides with the generating function for the Legendre 
polynomials~$P_n(\cosh(x))$, and, therefore,~$f_2(n,t)=n!\<P_n(\cosh(x))\>$. 
The average can now be completed with the aid of~\eqref{eq_F2_log}.
Multiplying it by~$|J(x)|\,P_n(\cosh(x))$, 
integrating by parts twice, and using the equation for the Legendre 
polynomials,~$P''_n(\mu)(\mu^2-1)+2\mu P'_n(\mu)=n(n+1)P_n(\mu)$, 
we~get:
\bea
\nonumber
{\dot f}_2(n,t) &=& (1+a)n(n+1)\,f_2(n,t)\virg{\rm whence}\\
f_2(n,t) &=& n!\exp\bigl\{n(n+1)(1+a)t\bigr\},
\eea
which is in agreement with~\exref{f_function_general}. 

As an example, consider moments of a magnetic field advected by 
the fluid. The contravariant vector in this case is~$B^i/\rho$, 
where~$\rho$ is the density of the fluid. Let us denote the moments 
of~$B^i$ as~$H_n =\<|{\bf B}|^{2n}\>$. Recalling that, in the contravariant 
case,~$g=1/\rho^2$, we get from~\exref{moments_universal}:
\bea
H_n= f_d(n,t)\bigl<g^{-n(d-1)/d}\bigr>^{\rm contra},
\eea
where for the $g$~average we use the formula~\exref{E_contra} in Eulerian 
frame, or~\exref{E_co} in Lagrangian frame. 

An analogous derivation can be carried out for a covariant vector, 
e.g., gradient of a passive scalar~$\nabla\theta$. 
For its moments~$C_n=\<|\nabla\theta|^{2n}\>$, we find:
\bea
C_n = f_d(n,t)\bigl<g^{n/d}\bigr>^{\rm co}, 
\eea
where for the $g$~average we use formulas~\exref{E_co} or~\exref{E_contra}
depending on the frame of reference.

\vskip5mm
\centerline{\bf On passively advected tensors}
\vskip3mm

We now briefly demonstrate how one can calculate exactly the moments of 
a passively advected higher-rank tensor~$\hat T$. Suppose that we are 
interested in some moment~$\langle{\hat T}^m\rangle$. After averaging 
with respect to the initial distribution of~$\hat T$, we are left with 
a combination of~${\rm Tr}^k({\hat g}^n)$, which are polynomials of 
degree~$nk$ in the eigenvalues of the metric~$\hat g$. But any symmetric 
polynomial of degree~$m$ can be expanded in Jack polynomials of degree~$m$, 
which can then be averaged exactly. The result will therefore be a linear 
combination of exponents growing at the rates given by~\exref{our_energies}.

For example, consider a contravariant bilinear form~$a^i b^k$, where~$a^i$ 
and~$b^k$ are initially independent Gaussian random vectors, 
$\<a^i_0 a^k_0\>=\<b^i_0 b^k_0\>=\delta^{ik}$, and find its second 
moment~$B_2=\<({\bf a}\cdot{\bf b})^2\>=\<{\rm Tr}({\hat g}^2)\>$. Then 
\bea
B_2 = \biggl<\sum_i^d \lambda_i^2\biggr> 
= \bigl<g^{2/d}\bigr>
\[\<J_{(2,0)}\> - {2\over3}\<J_{(1,1)}\>\right], 
\eea
where polynomials~$J_{(2,0)}$ and~$J_{(1,1)}$ are constructed 
in~\exref{polynomials_2}. The corresponding  eigenvalues 
are~$\tilde E^{(2)}_{(2,0)}=(d+4)(d-1)/d$ 
and~$\tilde E^{(2)}_{(1,1)}=(d^2-4)/d$, as follows from~\exref{our_energies}. 
The answer is:
\bea
\nonumber
B_2 = &{}&\exp \Biggl\{ \({8\over d^2}+{2\over d}\)\gamma t\Biggr\} \\
\nonumber
\times  \[ {d^2+2d\over3}\, \right. &{}& \left. \exp  
\Biggl\{ 2\tilde E^{(2)}_{(2,0)}(1+a)t\Biggr\} \right. \\
- \left. {d^2-d\over3}\, \right. &{}& \left. \exp  
\Biggl\{ 2\tilde E^{(2)}_{(1,1)}(1+a)t\Biggr\}\].
\eea

\vskip5mm

We are very grateful to Russell Kulsrud and Alexandre Polyakov for many 
important discussions. We would also like to thank John Krommes for useful 
comments. This work was supported by the U.~S. Department of Energy Contract 
No.~DE-AC02-76-CHO-3073. One of the authors~(SAB) was also supported 
by the Porter Ogden Jacobus Fellowship from Princeton University.

\appendix

\section{PDF of Eigenvalue Ratios}
\label{PDF_independent_variables}

The $\delta$-function in~\exref{Jacobian_gzeta} can  
be integrated over, and~$\zeta_1,\dots,\zeta_d$ reduced to 
$d-1$ {\em independent} variables, viz.~the logarithms of the eigenvalue 
ratios:~$x_n={1\over2}\log(\lambda_n/\lambda_d)={1\over2}(\zeta_n-\zeta_d)$. 
In these variables, the equation for~$F$ becomes:
\bea
\nonumber
\dt F = (1&+&a)\Biggl\{ \sum_{n=1}^{d-1}{\d^2 F\over\d x_n^2}
+ \sum_{n=1}^{d-1}{1\over\tanh(x_n)}\,{\d F\over\d x_n}\Biggr. \\
\nonumber
+{1\over2}\sum_{n\neq m}^{d-1}{\d^2 F\over\d x_n\d x_m}
&+&\left.{1\over4}\sum_{n\neq m}^{d-1}{1\over\sinh(x_n-x_m)}
\[{\sinh(x_n)\over\sinh(x_m)}{\d F\over\d x_n}\right.\right. \\
&-& \Biggl.\left. {\sinh(x_m)\over\sinh(x_n)}{\d F\over\d x_m}\]\Biggr\}.
\eea
The last two terms correspond to interactions between
different~$x$'s and only enter for $d\ge3$.
The normalization rule now is:
\bea
\nonumber
{2^{(d+2)(d-1)/2}\over d}&{}&
\int \prod\limits_{n<m}^{d-1}|\sinh(x_n-x_m)| \\
\label{Jacobian_x}
\times \prod\limits_{n=1}^{d-1}&{}&|\sinh(x_n)|{\rm d}x_n\,F = 1.
\eea

This form of the equation for~$F$ is most convenient for numerical 
solution and for geometric analysis such as that of~\secref{solutions}.

\section{Jack Polynomials}
\label{Jack_polynomials}

Jack polynomials~$J_{\mu}(x_1,\dots,x_d;\alpha)$ of degree~$m$ are 
homogeneous (of degree~$m$) polynomials, depending on~$d$ 
variables~$x_i$, and symmetric under all permutations of~$x_i$. 
They depend on a parameter~$\alpha$ and are labeled 
by partitions~$\mu$ of an integer number~$m$. 

Partition~$\mu$ of~$m$ is a non-increasing sequence of 
integers:~$\mu=(\mu_1\geq\dots\geq\mu_d)\in{\rm Z}^d_{\geq 0}$, 
such that~$m=\mu_1+\dots+\mu_d$. 
The polynomials~$J_{\mu}(x;\alpha)$ vanish if the number of parts~$l(\mu)$ 
is greater than the number of variables~$d$. 
Consider two partitions~$\mu$ and~$\lambda$ of the same length 
$l(\mu)=l(\lambda)=d$. One writes that~$\lambda \geq \mu$ 
if~$\lambda_1+\dots+\lambda_i \geq \mu_1+\dots+\mu_i$ for each~$i\leq d$. 
This defines the so-called {\em natural} (or {\em dominance}) ordering 
on partitions.

In order to give a formal definition of the Jack polinomials, first define 
the {\em monomial symmetric function}~$m_\mu$, corresponding to 
the partition~$\mu$: 
\bea
m_\mu=\sum x_1^{\mu_1}\cdots x_d^{\mu_d},
\eea
where the summation is over all permutations of~{$\mu_1,\dots,\mu_d$}. 
The Jack polynomials must, by definition, be represented as:
\bea
J_\lambda(x;\alpha) = \sum_{\mu\leq\lambda} u_{\lambda\mu}m_{\mu},
\label{first_condition}
\eea
and be the eigenfunctions of the Calogero-Sutherland Hamiltonian:
\bea
\label{Calogero_Sutherland}
H_S^{(\alpha)}=\sum_{i=1}^d\(x_i\,{\d\over\d x_i}\)^2 + 
{2\over\alpha}\sum_{i\neq j}^d{x_i^2\over x_i-x_j}\,{\d\over\d x_i}.
\eea
All coefficients~$u_{\lambda\mu}$ can be found recursively in terms 
of~$u_{\lambda\lambda}$ with the aid of this definition~\cite{Macdonald}.

Let us use this definition to construct the Jack polynomials 
for~$m=2$ and~$\alpha=2$. The corresponding partitions are~$(2,0,0,\dots)$ 
and~$(1,1,0,\dots)$. Using the first condition~\exref{first_condition}, 
we write:
\bea
\label{polynomials_2}
J_{(2,0)}(x;2) &=& \sum_{i=1}^d x^2_i + A\sum_{i<j}^d x_ix_j, \nonumber \\
J_{(1,1)}(x;2) &=& \sum_{i<j}^d x_ix_j.
\eea
The coefficient~$A$ must be found from the requirement that the polynomials 
be eigenfunctions of~\exref{Calogero_Sutherland}, which gives~$A=2/3$.

The eigenvalues (energies) corresponding to Jack polynomials are:
\bea
E_\mu^{(\alpha)}=\sum_{i=1}^d\mu_i^2 + {2\over\alpha}\sum_{i=1}^d(d-i)\,\mu_i.
\label{eigenvalue}
\eea
The energies~\exref{eigenvalue} depend on particular partitions~$\mu$. 
Any symmetric polynomial of degree~$m$ can be expanded in Jack polynomials 
of the same degree~$m$. Of all the other properties of the Jack polynomials 
we will need the following:
\bea
\prod_{i,j}^d{1\over(1-x_i y_j)^{1/\alpha}}
= \sum_\mu b_\mu(\alpha)J_\mu(x;\alpha)J_\mu(y;\alpha),
\label{Cauchy}
\eea
where the summation is performed over all possible partitions~$\mu$ 
of all non-negative integers, and~$b_{\mu}(\alpha)$ are some expansion 
coefficients that can be found in~\cite{Macdonald}. For our purposes 
we will need the formula~\exref{Cauchy} with the set~$\{y_j\}$ consisting 
of only one variable. In this case the expansion takes the form:
\bea
\prod_{i=1}^d{1\over(1-yx_i)^{1/\alpha}}
= \sum_{m=0}^\infty y^m Q_{(m)}(x;\alpha),
\label{Cauchy_1}
\eea
where~$\mu=(m)$ is a partition consisting of only one element. 
$Q_{(m)}(x;\alpha)$~stand for the properly normalized Jack polynomials. 
The explicit expression for~$Q_{(m)}(x;\alpha)$ is as follows:
\bea
Q_{(m)}(x;\alpha) = \sum_{1\leq i_1\dots\leq i_m}^d 
{(\theta)_{q_1}\cdots(\theta)_{q_d}\over q_1!\cdots q_d!}\,
x_{i_1}x_{i_2}\cdots x_{i_m},
\eea
where~$\theta=1/\alpha$,~$q_l=\#\{n|i_n=l\}$ is the multiplicity with which
the number~$l=1,2,\dots,d$ appears in~$i_1 \dots i_m$, 
and~$(\theta)_{q}=\theta(\theta+1)\cdots(\theta+q-1)$. 

To use these results we need to transform our Hamiltonian 
(in the square brackets in~\exref{master_eq_gzeta}) to
the form~\exref{Calogero_Sutherland}. Changing variables 
to~$\tilde\lambda_i=\exp(\zeta_i)=\lambda_i g^{-1/d}$, we get:
\bea
\nonumber
&{}&\tilde H_S=-{1\over d}\sum_{i,j}^d{\d^2\over{\d\zeta_i\d\zeta_j}} 
+ \sum_{i=1}^d{\d^2 \over\d\zeta_i^2}  \\
\nonumber
&+&\,{1\over2}\sum_{i<j}^d{1\over\tanh{1\over2}(\zeta_i-\zeta_j)} 
\({\d\over\d\zeta_i} - {\d\over\d\zeta_j}\) \\
\label{our_H}
= H_S^{(2)} 
&-& {1\over d}\(\sum_{i=1}^d\tilde\lambda_i{\d\over\d\tilde\lambda_i}\)^2 
- {d-1\over2}\sum_{i=1}^d\(\tilde\lambda_i{\d\over\d\tilde\lambda_i}\).
\eea
For any Jack polynomials of degree~$m$, the corresponding 
eigenvalues~$\tilde E_\mu^{(\alpha)}$ 
of the Hamiltonian~\exref{our_H} are:
\bea
\label{our_energies}
\tilde E_{\mu}^{(\alpha)} = 
E_{\mu}^{(\alpha)} - {m^2\over d} - {(d-1)\over 2}\,m .
\eea
In particular, the energy of~$Q_{(m)}(\tilde\lambda;2)$ is
\bea
E_m = {d-1\over d}\,m\(m + {d\over2}\).
\eea

We now notice that the averaged Jack polynomials~$Q_{(n)}(\tilde\lambda;2)$ 
and the functions~$f_d(n,t)/n!$ have the same generating function 
(see~\exref{generating_function} and~\exref{Cauchy_1}), whence 
\bea
f_d(n,t)=n!\bigl<Q_{(n)}(\tilde\lambda;2)\bigr>.
\eea
Since (\eqref{master_eq_gzeta})~$\dt F = 2(1+a)\tilde H_S F$, 
where $\tilde H_S$~is self-adjoint with respect to 
the measure~\exref{Jacobian_gzeta}, 
$f_d(n,t)$~satisfies: $\dot f_d(n,t) = 2(1+a)E_n f_d(n,t)$,
the solution of which (with correct initial condition) is 
the expression~\exref{f_function_general}.

\section{PDF for Contravariant Tensor}
\label{ap_contra}

The derivation of the main equations for the case of a contravariant 
tensor is quite similar to the case of the covariant tensor. 
Here we just explain the origin of the difference and write out 
the main results. The dynamical equation in the contravariant case reads:
\bea
\label{dynamic_g_contra}
\dt \phi^{ij} +
\xi^k\phi^{ij}_{,k} - \xi^i_{,k}\phi^{kj} - \xi^j_{,k}\phi^{ik} = 0.
\eea
The derivation of the master equation can be carried out the same 
way as in the covariant case and results in different coefficients 
in the $g$~part of~\eqref{master_eq_gzeta}. The~$\zeta$~part remains intact. 
This is not surprising, since the transition from~$\phi_{ij}$ 
to~$\phi^{ij}$ does not change the ratios of the eigenvalues, 
but results only in the inversion of the determinant:~$g\to1/g$.
Accordingly, the equation for the~PDF of a contravariant 
tensor,~${\tilde P}({\tilde g},\zeta)$, can be obtained  
from~\eqref{master_eq_gzeta} by 
substituting~$P={\tilde P}{\tilde g}^{d+1}$,~$g=1/{\tilde g}$. 
The resulting equation for the~PDF of the 
determinant,~${\tilde S}({\tilde g})$, is (dropping the overtildes):
\bea
\label{eq_S_contra}
\dt S = 2g^2{\d^2 S\over\d g^2} + (2d+3)\,g\,{\d S\over\d g} 
+ {1\over2}\,d\,(d+1)\,S, 
\eea
where time has been rescaled by the factor of~$\gamma$ as in~\secref{EL_PDF}.
When using this equation, we should remember that~$g$ now satisfies 
the same equation as~$1/\rho^2$, where~$\rho$ is the density of the medium.
\eqref{eq_S_contra} is written in the Eulerian frame. For
completeness, we write down the solution of \eqref{eq_S_contra} 
with initial distribution~$S(0,g)=\delta(g-1)$:
\bea
\label{S_contra}
S(t,g) = {g^{-(d+1)/2}\over\sqrt{8\pi\gamma t}} 
\exp\fl-{\bigl(\log(g)-\gamma t\bigr)^2\over8\gamma t}\fr.
\eea
The Lagrangian analogue of~\exref{S_contra} is obtained via multiplication 
by~$\rho=1/\sqrt{g}$.

\end{multicols}

\end{document}